# Theory of cargo and membrane trafficking

Lionel Foret[1], Lutz Brusch[2] and Frank Jülicher[3]

(1) Laboratoire de Physique Statistique, Ecole Normale Supérieure, UPMC, CNRS, 24 rue Lhomond, 75005 Paris, France. (2) Center for Information Services and High Performance Computing (ZIH), Technische Universität Dresden,  Nöthnitzer Straße 46, 01187 Dresden, Germany. (3) Max Planck Institute for the Physics of Complex Systems, Nöthnitzer Straße 38, 01187 Dresden, Germany.

Corresponding author: julicher@pks.mpg.de                    20 November 2014

## Abstract

Endocytosis underlies many cellular functions including signaling and nutrient uptake. The endocytosed cargo gets redistributed across a dynamic network of endosomes undergoing fusion and fission. Here, a theoretical approach is reviewed which can explain how the microscopic properties of endosome interactions cause the emergent macroscopic properties of cargo trafficking in the endosomal network. Predictions by the theory have been tested experimentally and include the inference of dependencies and parameter values of the microscopic processes. This theory could also be used to infer mechanisms of signal-trafficking crosstalk. It is applicable to in vivo systems since fixed samples at few time points suffice as input data.

## 1 Introduction

Any eukaryotic cell transports nutrients and signaling molecules, here collectively called cargo, between its plasma membrane and intracellular target compartments (Mellman (1996)). Receptor-mediated endocytosis, through inward budding of clathrin-coated vesicles (CCV) or clathrin-independent vesicles, provides the means for selective cargo uptake (Conner & Schmid (2003)). This endocytic cargo uptake inevitably is coupled to membrane and receptor uptake. Immediate delivery of the many small uptake vesicles to target compartments would risk a large membrane and receptor loss at the price of re-synthesis. A solution to this unfavorable consequence of coupled cargo and membrane trafficking is the intermediate accumulation of like cargo and recycling of membrane and a specific set of receptors to the plasma membrane (Sheff et al. (2002); Lauffenburger & Linderman (1993)). Early endosomes (EEs), i.e. several hundreds of intermediary vesicles per cell, fulfil this essential accumulation and sorting function (figure 1 A,C). EEs constitutively undergo homotypic fusion and fission among them and as a whole constitute the endosomal network of dynamically interlinked vesicles (Gruenberg & Maxfield (1995); Collinet et al. (2010)).

There is an intriguing question regarding the overall performance of the endosomal network. As the endosomal network is fuelled by small uptake vesicles and homotypic fusion and fission occur more or less randomly given heterogeneous cargo mixtures and their noisy perception, how could endosomes carrying large amounts of cargo be produced for more efficient cargo delivery to target compartments? In quantitative terms, the size distribution of the EEs should possess a broad tail as does a power law distribution (figure 1B,D,E).

The individual EEs operate independently of one another and have no information on the overall cargo distribution. Such a power law relation is characteristic for many complex systems (Ziemelis (2001)). Nontrivial collective properties can emerge at the scale of the whole endosomal network (here called the macroscopic level) that are not imposed explicitly at the level of the system components (the microscopic level). Repeated interactions among the components let the collective properties emerge or self-organize.

An additional tool that provides important insights is a theory of the endosomal network in which the details of single endosome interactions are described by mathematical terms and the collective effects of many such interactions can be analysed using mathematical methods. Once having linked microscopic to macroscopic properties through a theory then one can also use it as a means to infer microscopic properties that are not directly observable from readily-observable macroscopic properties. We review such a theory as developed by (Foret et al. (2012)) in the following section 2.1 and progress with its analysis for the particular case of low-density lipoprotein (LDL) trafficking in HeLa cells in section 2.2. The comparison to experimental data confirms the theoretical predictions and lets one estimate the microscopic parameter values. In subsequent sections, we describe selected mechanisms for regulating endosome interactions and discuss how they refine cargo and membrane trafficking.

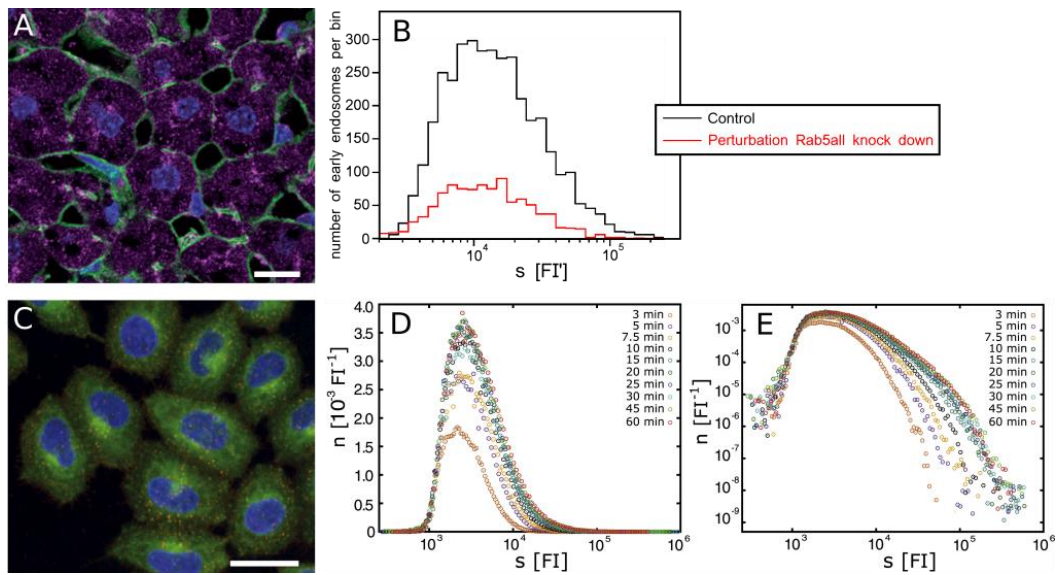

**Figure 1. Quantification of endosome distributions *in vivo* and *in vitro*. (A)** Fluorescence microscopy image of a mouse liver section stained for EEA1 (together with phalloidin (green) and DAPI (blue)), marking early endosomes (EEs) as magenta objects in hepatocytes *in vivo*. **(B)** Cargo distribution in early endosomes on lin-log scales (black curve for normal liver) with cargo amount *s* after 20min of internalizing labeled LDL. Perturbation of regulatory components of the endosomal network changes the cargo distribution *in vivo* (red curve for liver-specific Rab5 knock-down). **(C)** Fluorescence microscopy image of HeLa cells *in vitro* expressing GFP-Rab5 (green), loaded with labeled LDL cargo (red) and stained by DAPI (blue). **(D,E)** Cargo distribution in early endosomes on lin-log **(D)** and log-log **(E)** scales. The broad tail (with a power-law contribution) of the distribution is the hallmark of cargo accumulation through many rounds of homotypic EE-EE fusion and membrane recycling. Individual colors refer to different time points after adding labeled LDL cargo to the culture medium. FI and FI' are microscope-specific units of cargo fluorescence intensity. Scale bars are 20μm in (A) and 30μm in (C). Panels (A) and (B) are adapted by permission from Macmillan Publishers Ltd: Nature (Zeigerer et al. (2012)), copyright (2012). Panels (C) to (E) are reprinted from (Foret et al. (2012)), copyright (2012), with permission from Elsevier.

## 2 Theory of the endosomal network

### 2.1 General model

#### 2.1.1 Cargo distribution in a population of endosomes

At a given instant $t$, each endosome contains a certain amount of cargo (i.e. number of cargo molecules), see figure 1A and C. The overall population is characterized by the distribution of cargo defined as the number $n(s, t)$ of endosomes loaded with a number $s$ of cargo molecules. Repeated interactions between endosomes and with other cell organelles result in the continual redistribution of the cargo in the population and thus lead to the evolution of $n(s, t)$ over time. For example, when an endosome with 15 cargo molecules fuses with an endosome with 6 cargo molecules, these two endosomes disappear (then $n(15)$ and $n(6)$ decrease by one) and a new endosome with 21 cargo molecules appears ($n(21)$ increases by one). The profile of the distribution $n(s, t)$ and the way it evolves directly reflect the underlying dynamics of the many individual endosomes. The goal of the theory is to predict the emergent macroscopic behavior of $n(s, t)$ from the microscopic interaction processes at the single endosome level such as fusion and fission.

#### 2.1.2 Microscopic processes and parameters

Only six basic processes can modify the cargo distribution in the endosome population. They are sketched in figure 2 A-F and any other process would involve three or more simultaneously interacting endosomes, which is negligibly rare. The homotypic fusion of two endosomes loaded with the cargo amount $s$ and $s'$, respectively, leads to the disappearance of these two endosomes and to the formation of a new endosome containing $s + s'$ cargo (Gorvel et al. (1991); Gruenberg & Maxfield (1995); Rink et al. (2005); Rybin et al. (1996)), see panel C in figure 2. Homotypic endosome fission (D) is the reverse process: two new endosomes with the cargo amount s and s' arise from a single endosome containing $s + s'$ cargo (Bonifacino & Glick (2004); Rink et al. (2005); Skjeldal et al. (2012)). Beside these internal exchanges among individuals within the population, the endosome population receives and delivers cargo from and to other cell organelles by means of the four processes (A), (B), (D) and (E), but no other. Cell organelles are able to abruptly and irreversibly change their nature by a rapid modification of their biochemical composition. This process is called maturation, or sometime conversion. Non-endosomal structures can convert into endosomes and conversely endosomes can convert into organelles of another type. For example, endocytosed vesicles may convert into early endosomes (Conner & Schmid (2003); Gruenberg & Maxfield (1995)) while early endosomes can convert into late endosomes (Rink et al. (2005); Del Conte-Zerial et al. (2008); Poteryaev et al. (2010)). Conversion of a cargo-filled vesicle into an endosome is equivalent to the creation (biogenesis) of a new endosome with the same cargo amount as the original object (A). On the other hand, conversion of an endosome into another structure is equivalent to the removal of this endosome (and its cargo) from the population (F). Finally, individual endosomes can receive and release cargo by means of heterotypic fusion (B) and heterotypic fission (E), respectively. A heterotypic fusion event with a cargo-loaded vesicle results in the growth of the cargo amount in that endosome. The budding and scission of a vesicle or tubule decreases the cargo amount in the parent endosome but neither of these two processes changes the total number of endosomes. (A-F) is the exhaustive list of events able to affect the cargo distribution in an endosome network. However, depending on the type of endosomes and on the type of cargo, not necessarily all six processes are active.

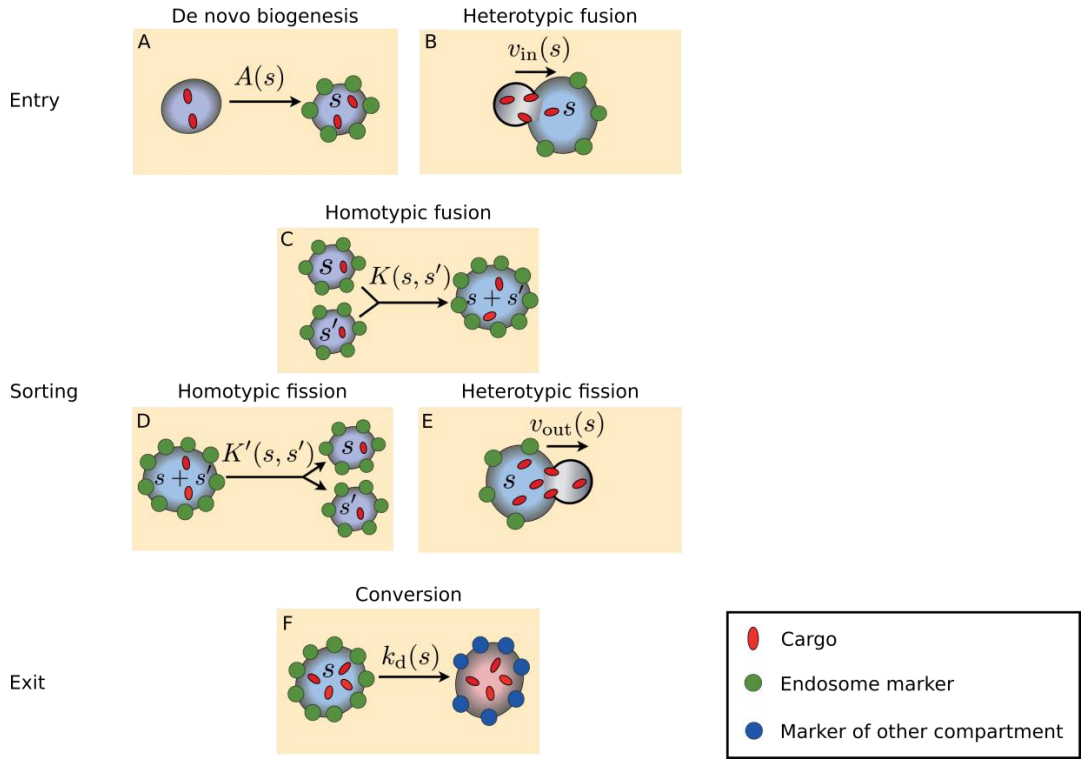

**Figure 2. There are six principle processes of endosome interactions.** Here these are ordered according to their sequence during cargo transport (red symbols). Green (blue) symbols denote membrane-associated proteins, such as Rab5 (Rab7), which confer functional specificity to endosomes. See section 2.1.2 for details on individual processes **(A)** to **(F)**. Reprinted from (Foret et al. (2012)), copyright (2012), with permission from Elsevier.

The six events (A-F) occur repeatedly in an apparently stochastic manner driving the evolution of the population. The probability that a fusion of two endosomes with cargo amount $s$ and $s'$ occurs during the short time interval $dt$ is $(s)n(s')dtK(s,s')$ ; it is proportional to the numbers of endosomes with $s$ and $s'$ cargo amount and to $dt$. The coefficient $K(s,s')$ is the fusion rate, whose value depends on the molecular details of the protein machineries responsible for endosome motility, tethering and membrane fusion. The theory at the endosome scale does implicitly include such details through the fusion rate $K(s,s')$. It is a function of $s$ and $s'$ since the cargo load could indirectly regulate the fusion machinery favoring or preventing fusion. In the same manner, we can define five more rates characterizing each of the processes A-F of figure 2. The probabilities associated to the occurrence, during $dt$, of the other events as ordered in figure 2 are $A(s)dt$, $v_{in}(s)n(s)dt$, $n(s+s')K'(s,s')dt$, $v_{out}(s)n(s)dt$ and $k_d(s)n(s)dt$ with $A(s)$ the *de novo* creation rate, $v_{in}(s)$ the cargo absorption rate, $K'(s,s')$ the fission rate, $v_{out}(s)$ the cargo release rate and $k_d(s)$ the conversion rate. The six rate functions are the context-specific input of the theory. Given these parameters, the goal is to calculate $n(s,t)$.

### 2.1.3 Governing equation

Having chosen the parameters (the six rates) and an initial state of the endosome population of a cell, one can simply simulate the stochastic evolution of $n(s,t)$ by repeatedly using the probabilities for each process to occur within short time intervals. Such a stochastic simulation can be efficiently performed using the Gillespie algorithm (Cao et al. (2004), Mukherji & O'Shea (2014)). Simulating the evolution several times and averaging, one can compute the evolution of the mean $n(s, t)$ that can be compared to experimental data.

Another approach is to derive a mathematical equation satisfied by the mean $n(s, t)$. Because of the large number of endosomes in a population, a kinetic equation similar to those used in chemical kinetics describes the evolution of the mean $n(s, t)$. It reads

$$\frac{\partial n(s)}{\partial t} = \frac{1}{2} \int_0^s K(s', s-s') n(s-s') n(s') ds' - \int_0^\infty K(s', s) n(s) n(s') ds'$$

$$+ \int_0^\infty K'(s, s') n(s+s') ds' - \frac{1}{2} \int_0^s K'(s-s', s') n(s) ds'$$

$$+ A(s) - k_d(s) n(s) - \frac{\partial}{\partial s} [v_{in}(s) n(s)] + \frac{\partial}{\partial s} [v_{out}(s) n(s)] \qquad (1)$$

Note that in order to make the mathematical analysis of the equation simpler, we have considered the cargo amount $s$ as a continuous variable. It follows that $n(s)$ is a number density, $n(s) ds$ being the mean number of endosomes with a cargo amount between $s$ and $s + ds$ for a vanishingly small $ds$. An analogous approach is used to describe coagulation of particles in physics (Smoluchowski (1916); Hogg et al. (1966)). Each term of the equation gives the average number of endosomes with $s$ cargo created or lost per time unit due to one of the six basic processes, with fusion and fission requiring two terms each. The merit of this approach is that exact or approximated mathematical solutions can be found for simple forms of the parameters. This allows us to gain mechanistic understanding of the contribution of each microscopic process to the macroscopic property $n(s, t)$ and on the way how it evolves.

## 2.2 Entry-fusion-exit model

Let's now apply the theory to predict the features of the distribution of cargo $n(s)$ for a simple set of parameters. Consider a hypothetical population in which new endosomes with small cargo amount $s_0$ appear at the rate $A = J/s_0$ (J is the cargo influx), fuse at the constant rate $K$ and convert at the constant rate $k_d$. Fissions (D and E) and heterotypic fusion (B) are not included (their rates are set to zero). The endosome population is initially devoid of cargo. We refer to this choice of parameters as the "entry-fusion-exit model" of endosome dynamics.

### 2.2.1 Theoretical predictions

In this simple model, the evolution of the total number $N = \int_0^\infty n(s) ds$ of endosomes in the population is governed by the equation,

$$\frac{dN}{dt} = \frac{J}{s_0} - \frac{KN^2}{2} - k_d N \ . \qquad (2)$$

From here on, we consider that fusion events are much more frequent than conversion events. Then the solution is approximately

$$N(t) = \sqrt{2J/Ks_0} \tanh\left(\sqrt{JK/2s_0} \, t\right) . \qquad (3)$$

At early times, while the number of endosomes is small, fusion (and conversion) is very rare, the number of endosomes rises as $N(t) = (J/s_0)t$. After some time, the number of fusion events and de novo creation events balance each other and the number of endosomes stays nearly constant at $N(t) = \sqrt{2J/Ks_0}$.

We now turn to the behavior of the full distribution $n(s,t)$ in the entry-fusion-exit model. Its evolution can be divided into three phases. In the first phase, the distribution $n(s,t)$ peaks at $s_0$ and its height $n_{max}$ behaves as the total number of endosomes, $n_{max}(t) = (b/s_0)N(t)$ with $b$ a numerical prefactor. It rises until fusion balances de novo creation. In the second phase, fusion events produce endosomes accumulating more and more cargo. The distribution $n(s,t)$ broadens more and more over time. In the third phase, endosome conversion curbs the growth of endosomes and yields a steady state: $n(s)$ does no longer evolve in phase 3. Mathematically, beyond the peak at $s_0$ and during the phases 2 and 3, the distribution is approximately

$$n(s) = \sqrt{J/2\pi K}\ s^{-3/2} \exp(-s/s^*(t)) \quad \text{with} \quad s^*(t) = \begin{cases} JKt^2 & \text{for } t \ll 1/k_d \\ 2JK/k_d^2 & \text{for } t \gg 1/k_d \end{cases} \tag{4}$$

The cargo distribution decays slowly as a power law $n_s(s) \sim s^{-3/2}$ up to a cargo value $s^*$ beyond which the decay is abrupt. This maximal cargo load $s^*$ increases with time during the second phase and finally saturates at a constant value in the third phase. This evolution is called "self-similar": as time progresses, $n(s,t)$ gets "stretched" along the $s$ axis until its final steady state profile.

These results illustrate how the mathematical analysis of the equation allows us to understand the effect of each basic process on the shape of $n(s,t)$. Comparing these predictions to experimental data, one finds that already the simple entry-fusion-exit model very accurately describes the trafficking of LDL cargo in HeLa cells.

### 2.2.2 Comparison of theory predictions to LDL trafficking in HeLa cells

Analyzing fluorescence microscopy images (labeling an early endosome marker and cargo), one can detect endosomes, measure for each of them their cumulative cargo fluorescence which corresponds to the cargo amount $s$ and then deduce the distribution $n(s)$, see figure 1. In an interesting experiment, HeLa cells stably expressing Rab5-GFP as early endosome marker were placed in a medium containing fluorescent LDL cargo. The distribution of LDL amounts in Rab5-endosomes was then measured at different times after the cells started to endocytose LDL. Figure 1E and F shows that the experimentally measured cargo distributions qualitatively validate the prediction by the entry-fusion-exit model: a peak at small s rises and saturates during the first 10 minutes (phase 1) and on a longer time scale, the distribution progressively broadens and apparently stretches along the s axis (phase 2), it finally reaches a stationary state after 30 minutes (phase 3). Beyond this qualitative agreement, the data reproduced in figure 1 bear all the mathematical features predicted in the preceding section. The evolutions of the height of the distribution's peak $n_{max}$ and of the total number of endosomes $N$ are very well fitted by the law (3), see figure 3A and C. In figure 3B, the distributions at different times where plotted on a rescaled axis with an $s^*$ chosen for each time such that the distribution tails overlap. Beyond the peak, all the curves fall onto the same curve, which is the function $x^{-3/2}\exp(-x)$ shown as solid line in figure 3B. This data collapse confirms the self-similar nature of the evolution and the accuracy of the predicted solution (4). Moreover $s^*(t)$ obtained by the curve rescaling evolves as predicted in equation (4): it first grows with time as $t^2$ and reaches a plateau after 30min (inset in figure 3B).

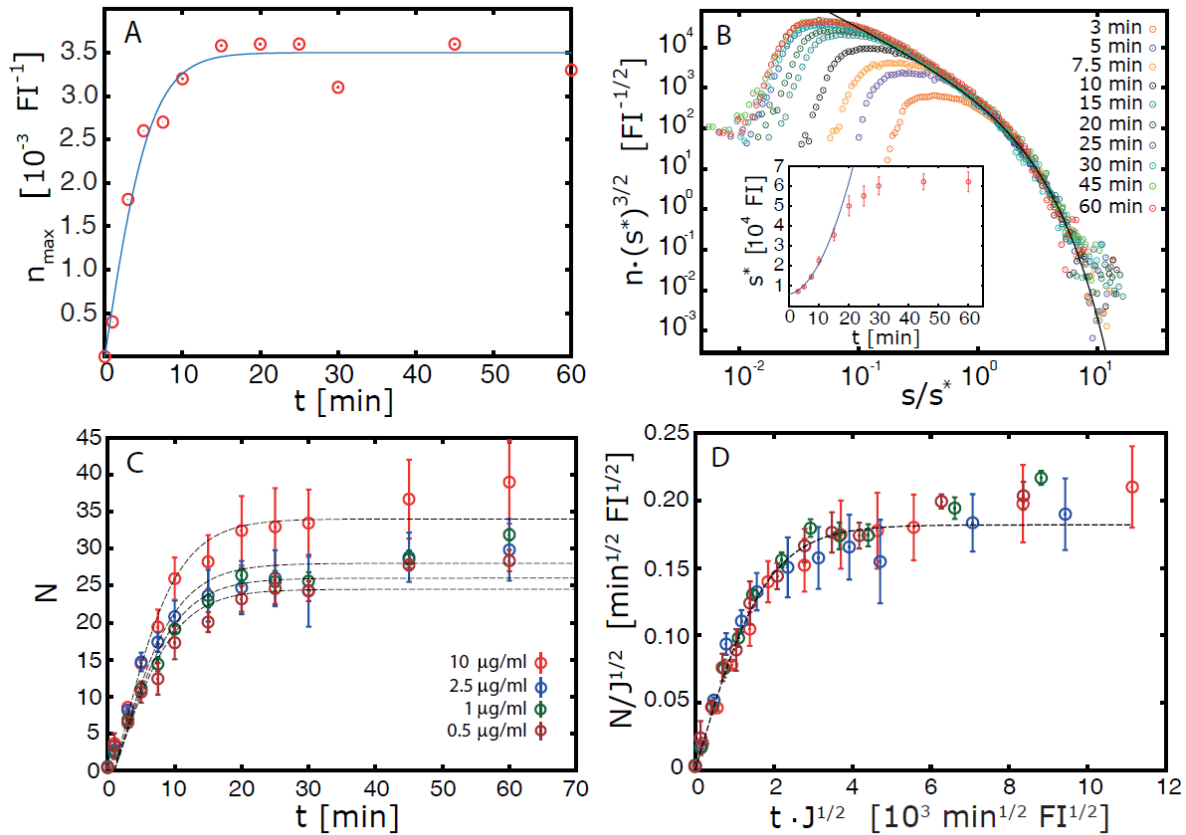

**Figure 3. Experimental verification of the theory of cargo trafficking in the endosomal network. (A)** Maximum values of the cargo distributions shown in figure 1D as function of time (symbols) verify the theory prediction (solid curve) by equation (3). **(B)** Rescaled cargo distributions (symbols) from figure 1E collapse on the predicted universal curve shape (equation (4)). The scaling parameter $s^*$ (inset) is correctly predicted by equation (4), the parabolic limit of the prediction for short times is shown as solid line, the saturation is also correctly predicted. **(C)** The dependence of the total number of cargo-filled early endosomes on time and medium composition (see legend) is correctly predicted (dashed curves) by equation (3). **(D)** Rescaling of the data from panel (C) with the predicted influx dependency collapses all data onto the universal time dependency as predicted by equation (3). Reprinted from (Foret et al. (2012)), copyright (2012), with permission from Elsevier.

In order to further test the predictions, one would like to experimentally tune some parameters. Changing the concentration of LDL in the cell culture medium results in the change of the LDL influx $J$ into each cell and thus into the early endosomal system. The value of $J$ can be easily deduced from the evolution of the total LDL fluorescence of the cell. The experiments were repeated using 4 different values of J. The evolution of $N(t)$ in the 4 cases is shown in figure 3C. When plotting $N/\sqrt{J}$ as a function of $\sqrt{J}t$, all the curves once again overlap as predicted by equation (3), figure 3D.

### 2.2.3 What do we learn?

Quantitative comparison between experimental data on the entire endosome population and the theoretical predictions allows us to reach unbiased conclusions regarding the organization of the early endosomal system. Degradative cargos such as LDL that enter the cell in many small structures are progressively concentrated in a few large endosomes by the mean of repeated random fusion and finally sent to the late endosomal system by conversion of early endosomes. Another scenario that one may propose after qualitative observations would lead to a markedly different profile and evolution for $n(s,t)$ and can thus be ruled out. Analogously, the data are only compatible with a constant fusion rate since otherwise the distribution features would be different, see (Foret et al. (2012)) for details. It indicates that LDL is a passive cargo weakly affecting the fusion machinery. A fit of the data yields the values of the microscopic parameters (entry rate, fusion rate, conversion rate) that are difficult to measure. The success of the entry-fusion-exit model does not mean that fission and heterotypic fusion do not exist but it means that they should have a small influence on the distribution of LDL cargo in early endosomes of HeLa cells.

## 3 Regulation of the microscopic processes

Cargo molecules change their receptors' activation state through binding and unbinding which itself is affected by cargo accumulation within individual endosomes and by gradual acidification of the endosomal lumen (Lauffenburger, D. A. & Linderman, J. J. (1993)). The receptors' cytoplasmic domains can transmit the binding signal to effector proteins with guanine nucleotide exchange (GEF) and GTPase-activation (GAP) activities (Tall et al. (2001)). Small GTPases of the Rab family act as master regulators of the endocytic protein machinery and can integrate the cargo-dependent GEF and GAP activities with other signals (Zerial & McBride (2001)). For EEs, the GTPase Rab5 plays this central role and increased Rab5 activity fosters cargo uptake (Zerial & McBride (2001)), homotypic EE fusion (Ohya et al. (2009)), homotypic EE fission (Zeigerer et al. (2012)) and, surprisingly, also EE to late endosome conversion (Rink et al. (2005); Del Conte-Zerial et al. (2008); Poteryaev et al. (2010)). The activation of the reciprocal microscopic processes fusion and fission by Rab5 conveys robustness to macroscopic properties such as total endosome number as demonstrated in an integrated theoretical and experimental study of Rab5 titration in hepatocytes (Zeigerer et al. (2012)). Different cargo types may affect Rab5 differently through their specific co-transported receptors. E.g. the cargoes EGF and LDL have been observed to specifically modulate the macroscopic properties of the endosomal network (Collinet et al. (2010)). In the theory presented above, the transport of different cargoes can be modelled by readjusting the parameter values for the rates of the microscopic processes. The regulatory functions of certain cargoes as well as possible dependencies of the microscopic processes on the physical size of the endosomes can in the theory be accounted for by the $s$-dependencies and additional time-dependencies of the six rate functions, see section 2.1.2.

## 4 Relation to kinetics of total cargo pool

The total cargo content $\Phi(t)$ per cell can be obtained from the cargo distribution $\Phi(t) = \int_0^\infty s\, n(s,t)ds$. The total cargo influx $J$ into the endosomal network is $J(t) = \int_0^\infty \big(s\, A(s,t) + v_{in}(s,t)n(s,t)\big)ds$. A general kinetic equation for $\Phi(t)$ can be derived from equation (1) by multiplication with $s$ and integration of both sides of the quality:

$$\frac{d\Phi}{dt} = J(t) - \int_0^\infty \big(k_d(s,t)s\, n(s,t) + v_{out}(s,t)n(s,t)\big)ds \ . \tag{5}$$

Only in special cases with $k_d(s,t) = const$ and $v_{out}(s,t) = k_{out} * s$, the general theory simplifies to the known kinetic equation $\frac{d\Phi}{dt} = J(t) - (k_d + k_{out})\Phi$ (Lauffenburger & Linderman (1993); Klausner et al. (1985); Tzafriri et al. (2004)). However, non-trivial system behavior including delayed forwarding of spikes of cargo or signals can only be accounted for by equation (5) which again requires to solve equation (1).

# 5 Implications for cargo sorting

Besides describing degradative cargo trafficking, this theory is also applicable to recycling cargo and membrane turnover itself. The heterotypic fission process (figure 2E) will then play the essential role. Both, homotypic fusion and heterotypic fission of membrane are tightly coupled as shall be pointed out at the end of this section. For arbitrary but qualitatively similar shapes of EEs before and after a fusion event, the volume grows as the third power of a linear dimension whereas the membrane area only grows with the second power leaving excess membrane that can be extracted through membrane tube formation and then be recycled towards the plasma membrane (Skjeldal et al. (2012)). The membrane tube separation from a post-fusion EE also facilitates the recycling of receptors and membrane-associated cargo. Cargo that is destined for accumulation and subsequent degradation (e.g. LDL) can be passively sorted from other cargo that is destined for recycling by separating the former cargo from the membrane. For instance, cargo taken up by receptor-mediated endocytosis can dissociate from its receptor upon the pH decrease in EEs compared to the extracellular space. A membrane tube has a particularly low volume to surface ratio and correspondingly little degradative cargo will leak into the recycling pathway. This passive sorting process without the need for receptors specifically localized to the bulk part of an EE is termed geometric sorting (Maxfield & McGraw  (2004)). Geometric sorting reaches high fidelity through multiple iterations of fusion and fission among subpopulations (e.g. first among EEs and subsequently among recycling endosomes).

# 6 Conclusion

Characteristic macroscopic properties of the endosomal network as a whole emerge from the many stochastic endosome interactions on the microscopic scale. One such macroscopic property of interest is the probability of finding a large concentrated cargo packet versus many small cargo packets within a cell. The underlying endosome interactions have been difficult to investigate due to their highly parallel and stochastic occurrence. Here, a unifying theory has been presented which links the microscopic properties of endosome interactions to the emergent macroscopic properties of cargo trafficking in the endosomal network. The theory allows to infer microscopic kinetic parameters such as the fusion rate between endosomes from the statistics of the whole endosomal network at fixed time points after cargo internalization, lending this approach to in vivo studies that require fixing of samples. The results are robust and sensitive because the cargo statistics average over many stochastic events in many cells.